May 19, 2015

# Biomechanical and Structural features of CS2 fimbriae of Enterotoxigenic *Escherichia coli*

Narges Mortezaei[a], Bhupender Singh[a,b], Johan Zakrisson[a], Esther Bullitt[c] and Magnus Andersson[a,b*]

[a]Department of Physics, Umeå University, [b]Umeå Centre for Microbial Research (UCMR), Umeå University SE-901 87 Umeå, Sweden, [c]Department of Physiology and Biophysics, Boston University School of Medicine, Boston, MA 02118, USA

**Running title:** The biomechanical and structural properties of CS2 fimbriae

## Abstract

Enterotoxigenic *Escherichia coli* (ETEC) are a major cause of diarrhea worldwide, and infection of children in underdeveloped countries often leads to high mortality rates. Isolated ETEC express a plethora of colonization factors (fimbriae/pili), of which CFA/I and CFA/II that are assembled via the alternate chaperone pathway (ACP), are amongst the most common. Fimbriae are filamentous structures, whose shafts are primarily composed of helically arranged single pilin-protein subunits, with a unique biomechanical capability allowing them to unwind and rewind. A sustained ETEC infection, under adverse conditions of dynamic shear forces, is primarily attributed to this biomechanical feature of ETEC fimbriae. Recent understandings about the role of fimbriae as virulence factors are pointing to an evolutionary adaptation of their structural and biomechanical features. In this work, we investigated the biophysical properties of CS2 fimbriae from the CFA/II group. Homology modelling its major structural subunit CotA reveals structural clues and these are related to the niche in which they are expressed. Using optical tweezers force spectroscopy we found that CS2 fimbriae unwind at a constant force of 10 pN and have a corner velocity of 1300 nm/s, i.e., the velocity at which the force required for unwinding rises exponentially with increased speed. The biophysical properties of CS2 fimbriae assessed in this work classify them into a low-force unwinding group of fimbriae together with the CFA/I and CS20 fimbriae expressed by ETEC strains. The three fimbriae are expressed by ETEC, colonize in similar gut environments, and exhibit similar biophysical features, but differ in their biogenesis. Our observation suggests that the environment has a strong impact on the biophysical characteristics of fimbriae expressed by ETEC.

## Introduction

Enterotoxigenic *Escherichia coli* (ETEC) diarrheal infection is considered a prevailing health problem in developing countries since it is one of the major causes of deaths among infants and children. Additionally, ETEC infection is the leading cause of traveler's diarrhea with >60% of visitors to these countries experiencing diarrhea, which in some cases can even trigger an irritable bowel syndrome (1, 2). Pathogenesis of ETEC infections relies on bacterial attachment, via specialized fimbriae organelles, to the host intestine leading to subsequent release of either heat-labile (LT) or heat-stable (ST) enterotoxins (3–6).

Various strains of human ETEC express numerous serologically distinct fimbriae that are assembled either via the alternate (ACP) or classical (CUP) chaperone usher pathways (7). Colonization factor antigens - CFA/I and CFA/II fimbriae belong to the ACP family or class 5 group, of which CFA/I fimbriae is an archetype and the most extensively studied member (8–11). The CFA/II group, which comprises three different coli surface antigens – CS1, CS2 and CS3, shares a common assembly pathway but differs from that of CFA/I fimbriae in hemagglutination properties (3, 10, 12, 13). The CFA/II group fimbriae have been known for their role in causing diarrhea for almost 50 years, however little of their structural and biomechanical features are yet known (14–16). For example, micrographs of the three fimbriae indicate that CS1 and CS2 are wider and more rigid than CS3, and CS1 is a helical filament (17–20). However, whether or not CS2 or CS3 are helix-like structures is unknown.

Sustained adhesion of both bacteria and eukaryotic cells can be facilitated by a reduction in force on their tethers. Membrane tethers are formed by some cell types to maintain cellular adhesion by reducing the load on receptor-ligand complexes. For example, a neutrophil that is attached to an inflamed endothelium cell is exposed to a drag force. The cell extends long membrane tethers, partially reducing the load experienced by the P-selection and PSGL-1 bond, thereby increasing the lifetime of the complex (21). Similarly, unwinding is believed to be an important biomechanical property of fimbriae, to facilitate sustained adhesion of bacteria to their target cell (22, 23). Fimbria unwinding lowers the force on the adhesin and receptor bond, thereby reducing the probability of bacteria detachment. Previous data have shown that CFA/I pilins assemble into helix-like fimbriae that are easy to unwind in comparison to e.g., the uropathogenic *Escherichia coli* (UPEC) expressed P and type 1 fimbriae (24). Further similarities between fimbrial unwinding and tether formation by other cell types are detailed and explained more fully in ref. (25).

Interestingly, fimbriae expressed in a specific microenvironment exhibit similar properties, i.e., the unwinding force of CS20 is more similar to CFA/I than to P fimbriae even though CS20 share higher amino acid identity with the pilins of P fimbriae (26). Thus, the structural and biomechanical characteristics of fimbriae appear to play a determinant role in *E. coli* colonization and pathogenesis in a specific organ, i.e.,

*E. coli* strains sharing a common niche also express fimbriae with similar bio-physical properties (26). ETEC expressing CS2 fimbriae are known to localize to the small intestine, an environment similar to that of ETEC expressing CFA/I fimbriae of the ACP family and CS20 fimbriae of the CUP. We therefore hypothesized that CS2 would exhibit structural and biophysical similarities to those of CFA/I and CS20 fimbriae.

In this work, we examined this hypothesis by investigating the biophysical strength of the CS2 fimbriae of the CFA/II group using sub-pN force-spectroscopy instrumentation. Since a crystal structure of the major pilin subunit is lacking, we elucidated the structure of the major pilin, CotA, using homology modeling. Our data show that the CS2 fimbriae unwind at a constant force of 10 pN, demonstrating that the macromolecular structure of CS2 is helical with weak layer-to-layer (*ll*) interactions. While a three-dimensional structure of the CS2 shaft is not available, we have used a homology model of CotA and force spectroscopy results to reveal interesting biophysical features of CS2 fimbriae. Our data demonstrate similarities between this CFA/II group fimbria and fimbriae from the CFA/I group. The results in this work, together with previous data from the literature, place ACP and CUP ETEC-expressed adhesion organelles into a low-force unwinding group.

## Material and Methods

### Bacterial strains and growth condition

The C91F strain of enterotoxigenic *Escherichia coli* expressing CS2 fimbriae was used in this study. For expression of CS2 fimbriae, C91F strain was grown on CFA plates at 37 ˚C for overnight, and were re-streaked once again on CFA plate and grown overnight at 37˚C before analysis. Expression of CS2 fimbriae was confirmed by Atomic Force Microscopy (AFM) (see Fig. 1).

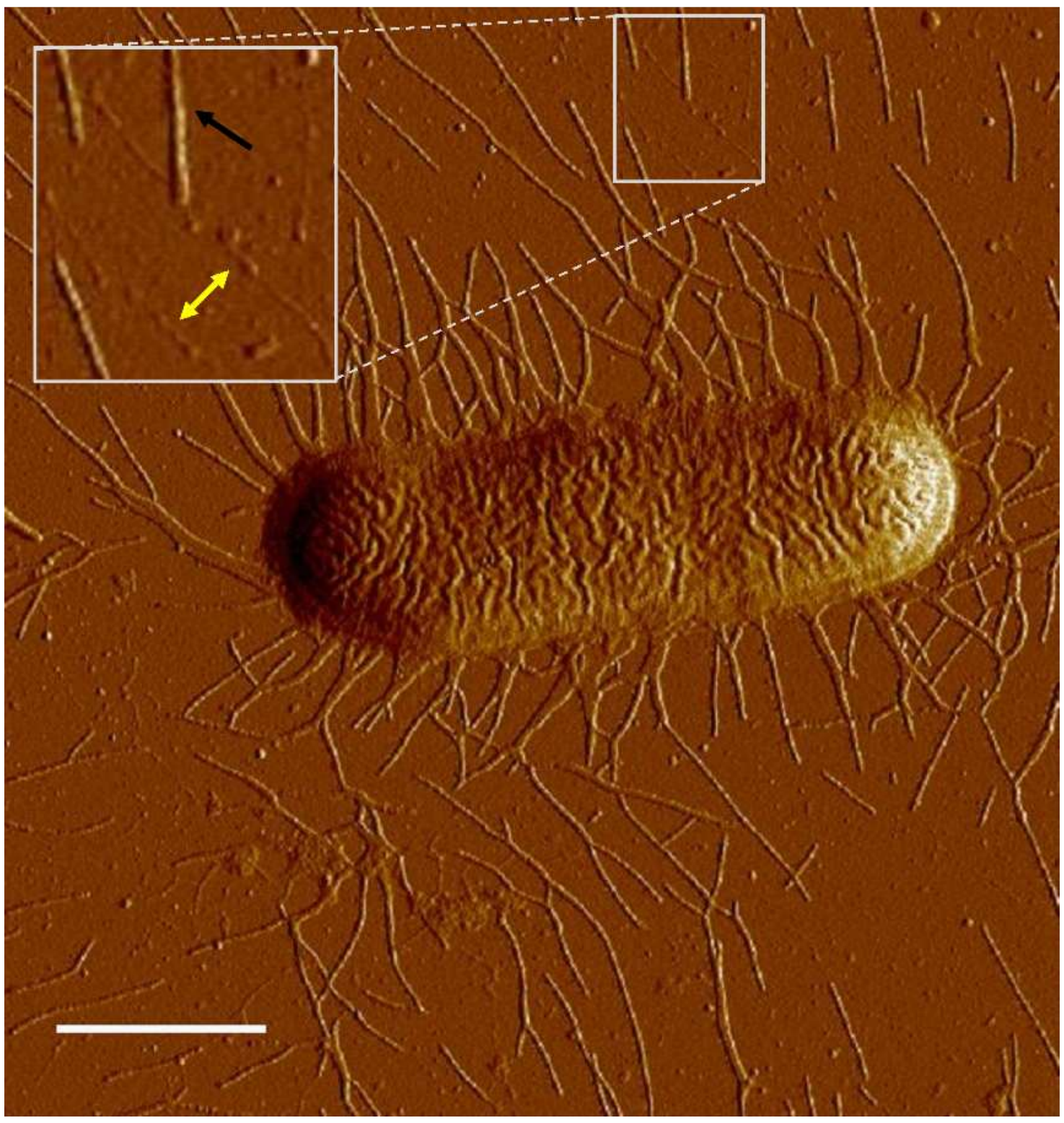

**Figure 1. Atomic Force Microscopy of a C91F ETEC bacterium expressing CS2 fimbriae.** The AFM micrograph shows a single C91F bacterium expressing CS2 fimbriae. The black arrow points at a fimbria in its helical state whereas the yellow arrows point at unwound fimbriae. Scale bar is 1.0 $\mu$m.

## Sample preparation and force spectroscopy measurement

Force spectroscopy experiments were performed using optical tweezers with sub-pN sensitivity. The setup of the instrumentation and assay is described in detail in references (27, 28). Briefly, the OT setup was constructed around an inverted microscope (Olympus IX71, Olympus) with a high numerical aperture oil immersion objective (model: UplanFl 100X N.A. = 1.35, Olympus). A Nd:$YVO_4$ laser (model: Millennia IR, Spectra Physics) that operates at 1064 nm in CW mode and run with an output power of 1.0 W was used for trapping. The position of a trapped bead and thereby the force, was monitored by projecting the beam of a low power fiber-coupled HeNe-laser (operates at 632.8 nm) onto a position-sensitive detector (L20 SU9, Sitek Electro Optics).

Bacteria expressing CS2 fimbriae were resuspended in Phosphate Buffer Saline (PBS 1x, pH7.4) and diluted in PBS 1:100 to a suitable concentration (maximum of 10 bacteria in the field of view) for experiments. Prior to making the flow chamber a 1:500 suspension of 9.5 $\mu$m CML latex beads (product no.2-10000, Interfacial Dynamics Corp.) in Milli-Q water was prepared. These larger sized beads were used to mount bacteria away from the coverslips to avoid any interactions with the surface. 10 $\mu$l of the bead-water suspension was dropped on 24×60 mm coverslips (no.1, Knittel Glass) and put into the oven for 60 minutes at 60 °C to immobilize the beads to the surface (10). To be able to adhere bacteria to the beads a solution of 20 $\mu$l of 0.01% Poly-L-Lysine (Catalogue #P4832, Sigma-Aldrich) was added to the coverslips, which after 45 minutes incubation at 60°C, were stored in a dust free box until use. A ring of vacuum grease (DOW CORNING®) was added around the area containing the Poly-L-Lysine coated beads on one of the coverslips. Gently, a 3 $\mu$l suspension of bacteria and a 3 $\mu$l suspension of probe beads (Surfactant-Free 2.5 $\mu$m White Amidine Polystyrene Latex beads, product No. 3-2600, Invitrogen) was dropped onto the area, which was then sealed by placing a 20×20 mm coverslip (no.1, Knittel Glass) on top.

An experiment was performed by trapping a single bacterium, at low laser power to make sure that the bacterium was not harmed, and firmly mounting it onto a Poly-L-lysine coated 10 $\mu$m latex bead as described in (29). Subsequently, a 2 $\mu$m amidine bead was trapped and the stiffness of the trap was calibrated using the power spectrum method (30). The stability of the setup as well as the optimal calibration time was measured using the Allan variance method for optical tweezers given in ref. (31). The trapped bead was thereafter attached to a fimbria by moving the bead in proximity to the mounted bacterium, but making sure that the bead was sufficiently far away from the bacterium to prevent a multitude of fimbriae attaching. With a fimbria attached to the bead (sometimes 2-5 fimbriae) the piezo-stage (PI-P5613CD, Physik Instruments) was translated using an in-house designed LabView program.

## Atomic force microscopy

Bacteria expressing CS2 were resuspended in 50 $\mu$l of filtered Milli-Q water. 10 $\mu$l of suspension was then placed onto freshly cleaved ruby red mica sheet (Goodfellow Cmbridge Ltd., Cambridge). The cells were incubated for 5 min at room temperature prior to being placed into a desiccator for ~ 2 h. Images were then collected with a Nanoscope V Multimode8 AFM setup (Bruker software) using Bruker ScanAsyst mode with Bruker ScanAsyst-air probe oscillated at a resonant frequency of 50-90 kHz (32).

## The Sticky-chain model for helix-like biopolymers

The unwinding velocity, $\dot{L}$, of helix-like biopolymer under tensile force, $F$, can be described using the opening and closing rates of the layer-to-layer interactions, $k_{AB}(F)$ and $k_{BA}(F)$, and the opening length $\Delta x_{AB}$ of a subunit,

$$\dot{L} = \left[k_{AB}(F) - k_{BA}(F)\right]\Delta x_{AB}, \tag{1}$$

where A and B represent the opened and closed states respectively (33, 34). Using the nomenclature defined by Zakrisson et al. (22), the unwinding velocity can be rewritten as,

$$\dot{L} = \Delta x_{\mathrm{AB}} k_{\mathrm{AB}}^{th} \left( e^{F \Delta x_{\mathrm{AT}} \beta} - e^{(V_0 - F \Delta x_{\mathrm{TB}})\beta} \right), \tag{2}$$

where $k_{\mathrm{AB}}^{th}$ is the thermal bond opening rate, $\beta = 1/kT$, where $k$ the Boltzmann's constant, $T$ is the temperature, $V_0$ is the energy difference between the ground and transition state, and $\Delta x_{\mathrm{TB}}$ is the distance between the transition state and the open state. From this expression, it is possible to denote the corner velocity, $\dot{L}^*$, as the highest extension velocity that can be used without the need to include the dynamic behavior of the polymer, i.e., for low extension velocities the opening and closing rates are in balance whereas for high extension velocities the closing rate can be neglected. This gives the following expression,

$$\dot{L}^* = \Delta x_{\mathrm{AB}} k_{\mathrm{AB}}^{th} e^{V_0 \Delta x_{\mathrm{AT}} \beta / \Delta x_{\mathrm{AB}}}. \tag{3}$$

By combining Eq. (2) and (3), the unwinding velocity can finally be expressed using the corner velocity,

$$\dot{L} = \dot{L}^* e^{(F - F_{SS}) \Delta x_{\mathrm{AT}} \beta} \left[ 1 - e^{-(F - F_{SS}) \Delta x_{\mathrm{AB}} \beta} \right]. \tag{4}$$

# Results

## Atomic force imaging of CS2 fimbriae

We used Atomic Force Microscopy to image CS2 fimbriae at high magnification. A representative micrograph of a single cell grown under normal conditions is shown in Fig. 1. The average length of fimbriae was measured from AFM micrographs such as shown in Fig. S1, using ImageJ software. CS2 fimbriae were arranged peritrichously on the bacterial cell surface, and had a length of 0.88 ± 0.34 $\mu$m (the number of samples were $n$ = 270). These fimbriae were found in two morphologies, where most fimbriae attached to the cell there were wider structures (black arrow) and only a few were narrower structures (yellow arrow). This suggested that the wider structures had intact quaternary structure, indicating that *ll* interactions of CS2 fimbriae are strong enough to be maintained during the sample preparation. However, the few narrower structures identified in the micrographs suggested that CS2 fimbriae can be unwound. Thus, these findings called for an investigation of the properties of the major pilin subunit.

## Homology modelling of the major structural subunit, CotA using CfaB as a template

CS2 fimbriae share a close genetic and pathological relationship with CFA/I fimbriae of the class 5 family. Each CS2 fimbria consists of continuously repeating major pilin subunits (CotA) and a tip-localized minor subunit (CotD). The mature CotA protein is composed of 147 amino acids, and shows significant homology to CfaB, the major pilin subunit of CFA/I fimbriae (PDB ID: 3F84) (23). The sequence alignment, which is presented in Fig. S2, was performed using the Clustal Omega algorithm (35); these two homologous pilin subunits share 51% identity and 80% similarity. Utilizing this high similarity, the CotA structure was homology modeled using the MODELLER software (36). Figure 2 shows the model of the outer (upper) and the inner (lower) representation of the CotA subunit as a ribbon structure and as surface views of both the hydrophobicity and the charge distribution, respectively. The homology model of the CotA was superimposed with the CfaB template and the root mean square difference between the backbone C-α atoms was assessed to 0.37 Å. Moreover, the surface properties of CotA and CfaB, the hydrophobicity and electrostatic potential, were thereafter examined using UCSF Chimera software (37). To investigate the surface potential of the model and the template we colored the negative and positive residues red and blue, respectively, as seen in the surface views of the charge distributions of two faces for both CfaB and CotA presented side-by-side in Fig. S3. In the upper panel the inner surface of the hydrophobic groove (tan) is visible running approximately vertically, in the center of the subunit. According to the charge distribution the CotA is slightly more negatively charged than the CfaB.

Residues in both the model and the template were also selected and shown according to their hydrophobicity and hydrophilicity, as shown in Fig. S4. Since the structural model of a pilus filament has not yet been determined we are not able to make an optimal fit of a subunit's position and exactly determine the inner and outer surfaces. However, by assuming that CotA and CfaB subunits have similar orientation it is possible to estimate the inner and outer surfaces. Using this assumption the model indicates that a majority of the area of the outer surfaces are hydrophilic whereas the inner surfaces are hydrophobic, which is in accordance with other helix-like adhesion fimbriae. Finally, the deep hydrophobic groove of the subunit is clearly visible in the center of the two structures.

With information of both the CS2 macromolecular structure, assessed from static AFM images, and information of the major pilin subunit using homology modeling, the next step was to gain information of the dynamic properties via force spectroscopy measurements.

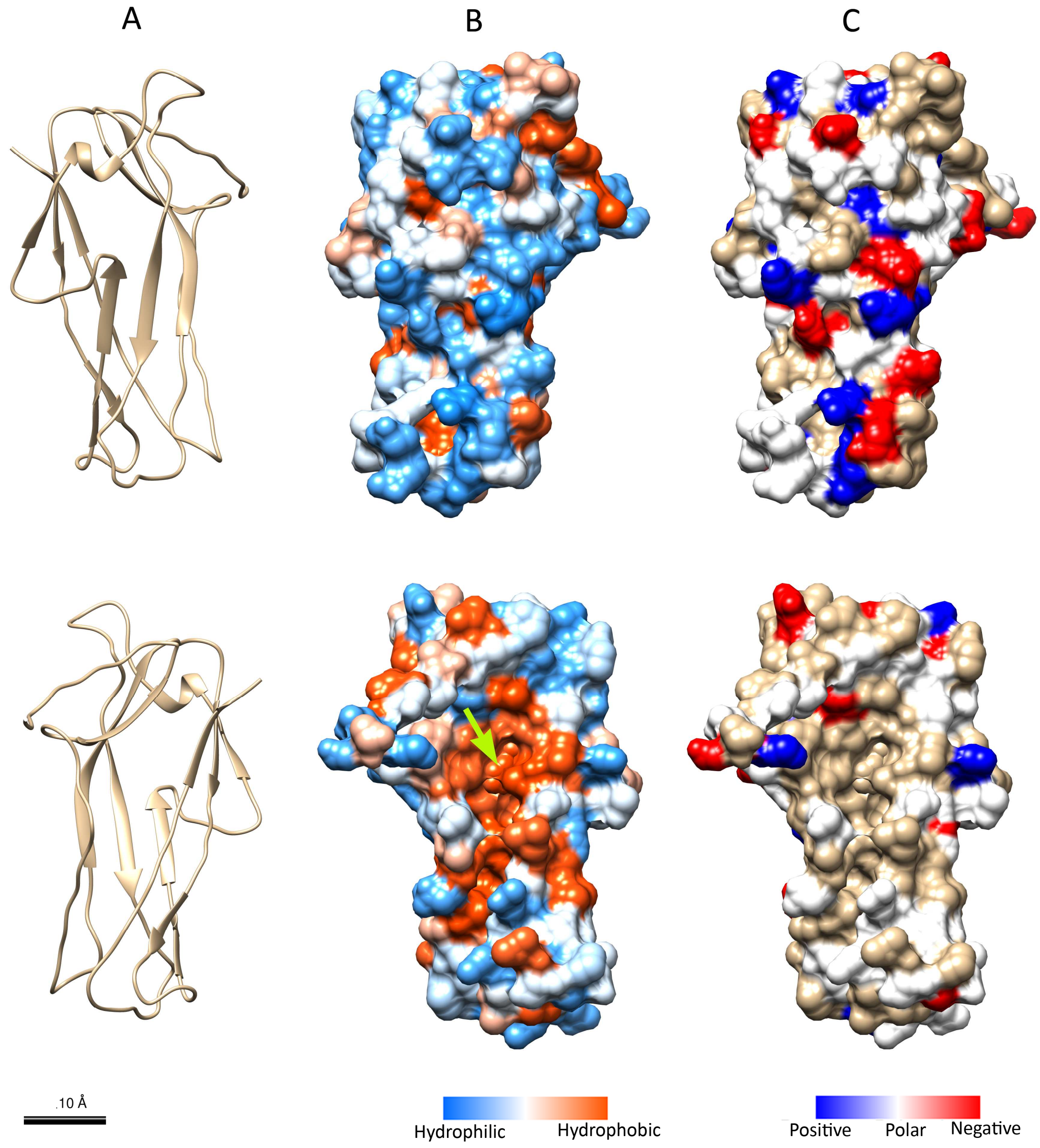

**Figure 2. Homology modeling the major structural subunit CotA.** The upper and lower panels show the model with its outer and inner surfaces, respectively. The panels represent the model as; A) a ribbon structure, B) a surface hydrophobicity map, and C) a surface charge map. The hydrophobic groove is clearly visible (green arrow) in the central part of the lower figure B. Scale bar for all panels is 10 Å.

## CS2-fimbrial response to tensile force

Mechanical response of the CS2 fimbriae to force was measured using optical tweezers force spectroscopy under steady-state conditions (38). According to our homology modelling data, CotA showed a negative charge distribution on the outer surface suggesting the possibility of a strong non-specific interaction with positively charged amidine microspheres. Our test run confirmed a strong enough interaction between the two, therefore, we used amidine microspheres for our force spectroscopy measurements. A representative force curve of a CS2 fimbria with three clearly distinct regions is presented in Fig. 3. CS2 first responds to tensile stress by a linearly increasing force, i.e., stretching layers gives a similar response as a Hookean spring and we denote this as region-I. After reaching a threshold in the structural resistance of the layers a transient change to a constant force response at ~10 pN can be seen; this is denoted region-II, and originates from unwinding of the individual shaft subunits. We marked this region by two dashed blue lines in the representative data shown in Fig. 3. The average unwinding force was calculated from 90 independent measurements and assessed to 10 ± 1.5 pN. After complete unwinding the force increases linearly with fimbria extension to ~7.8 $\mu$m, shown as region-III. After complete unwinding the force increases linearly with fimbria extension to ~7.8 $\mu$m, shown as region-III. As seen in the re-winding curve (blue) in Figure 4, a drop in force, ~6.9-6.8 $\mu$m, is required to nucleate re-coiling of the helical filament. The need for nucleation to begin helical rewinding supports a model in which the helical region is fully unwound prior to entering region III, and this region can be attributed to stretching of subunits already aligned in an open coil.

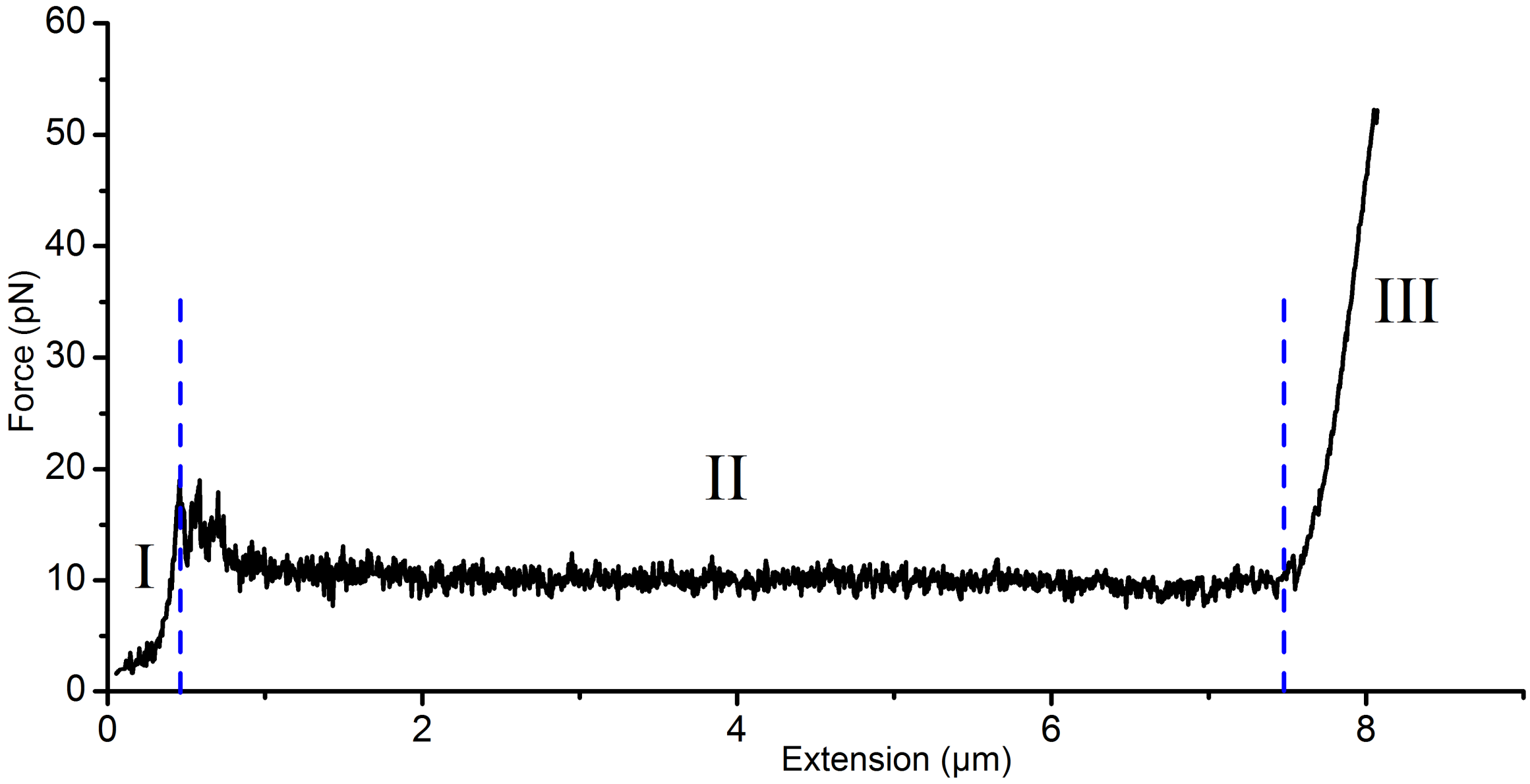

**Figure 3. Force spectroscopy measurement of a CS2 fimbria.** The black curve represents the unwinding force response of a CS2 fimbria at a velocity of 0.1 $\mu$m/s. The unwinding force response shows three distinct regions: region-I a linear increase of the force; region-II a constant force plateau (region between the blue dashed lines); and region-III. Initially the curve shows force peaks that originate from weakly attached fimbriae that detached with extension.

We observed that the force response of multiple fimbriae was additive as seen for other helix-like fimbriae (39). Figure 4 shows the extension of three fimbriae of different lengths. Similar to that of a single fimbria, the force response first increases linearly with extension, however with an increased stiffness, since three fimbriae were attached to the bead and the force was equally shared between these fimbriae. The transition to the constant force plateaus throughout the measurement took place for even multiples of a single fimbria. That is, first all three fimbriae unwound giving rise to a plateau force of 30 pN; at an extension of ~4.5 $\mu$m, the shortest fimbria was fully unwound and entered region-III before it detached from the bead. The two remaining fimbriae continued to unwind at ~20 pN for ~1 $\mu$m until the second fimbria reached region-III and detached. Eventually, the longest fimbria unwound an additional ~0.5 $\mu$m until it entered the region-III. At this point, we reversed the motion of the stage and allowed the fimbria to rewind (blue curve, Fig. 4). The rewinding force curve continued along the unwinding curve except for a 5 pN force drop that occurred between 6.8 and 6.6 $\mu$m, which originated from the lack of a nucleation kernel that is required for regenerating the helical form of the fimbrial shaft after complete unwinding. An additional force measurement that showed multiple fimbrial binding in the first sequence with four consecutive unwinding and rewinding sequences is presented in Fig. S5.

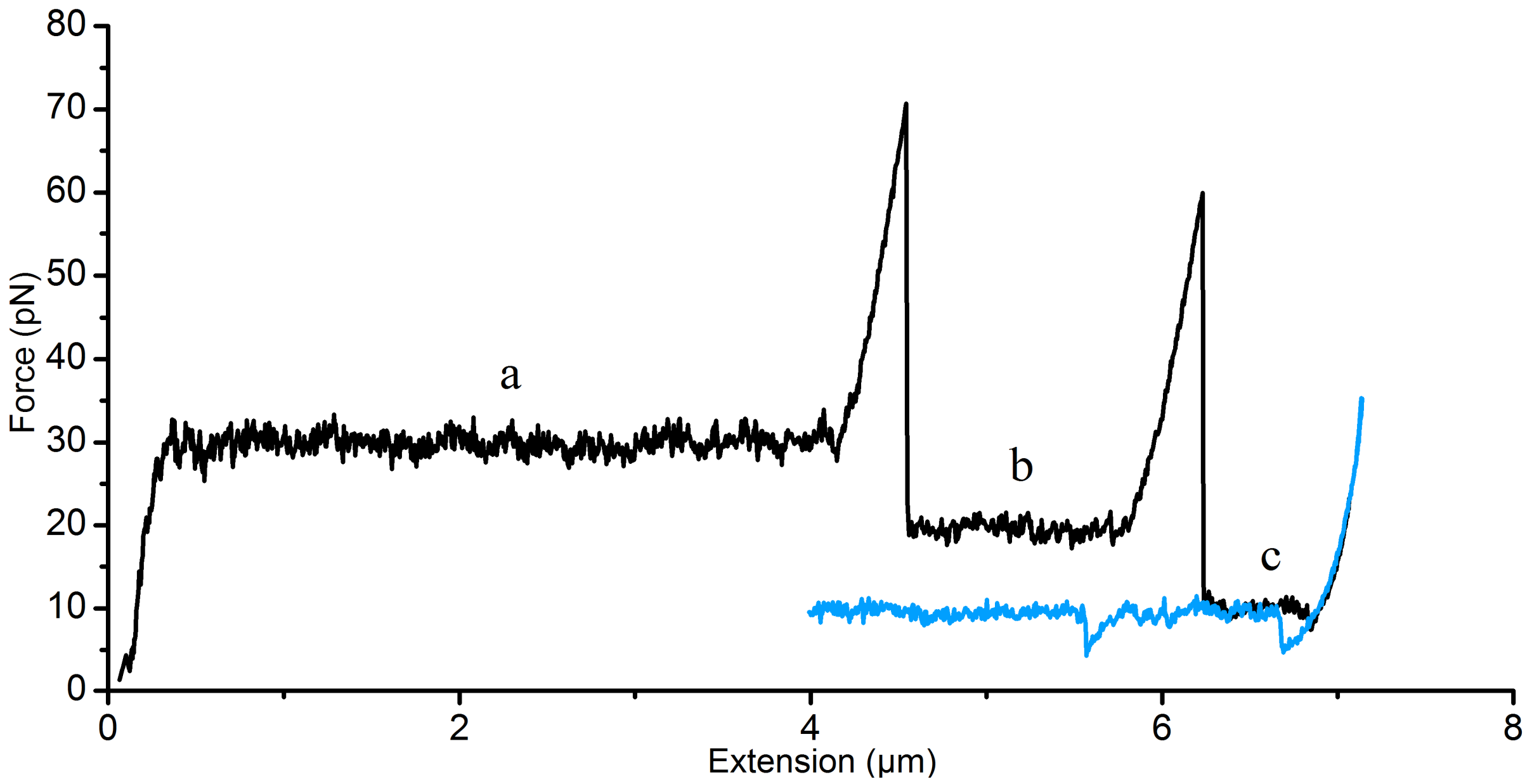

**Figure 4. Force spectroscopy measurement of multiple CS2 fimbriae.** The unwinding (black) and rewinding (blue) response of three simultaneously attached fimbriae where in; (a) all three fimbriae are simultaneously extended until the shortest fimbria detaches at ~4.3 $\mu$m resulting in the force drop to 20 pN; (b) the two remaining fimbriae are further extended until the shorter detached at ~6 $\mu$m with a corresponding drop in force to 10 pN; (c) the response of a single fimbria. The blue curve shows the corresponding rewinding response of the single fimbria.

## Dynamic force responses of CS2

To investigate the dynamic response of CS2 fimbriae we applied dynamic force spectroscopy (DFS) using optical tweezers (28). Prior to a DFS measurement a fimbria was slowly unwound to measure the total length of the unwinding region-II and to remove any possibilities of multi-fimbrial bindings to the bead. With the known length of region-II the fimbria was allowed to partially rewind to make sure that a 2 $\mu$m long region could be unwound without reaching region-III. The 2 $\mu$m long region was thereafter unwound at five different velocities, i.e., 0.1, 0.4, 1.6, 6.4, 25.6 $\mu$m/s and the corresponding position and force response was sampled at 5 kHz. Between each pull, the fimbria was allowed to rewind under steady-state conditions (0.1 $\mu$m/s) and a short pause of 2 s was introduced before the next pull. An example of a DFS measurement of a CS2 fimbria is shown in Fig. S6, with raw data presented by the grey curves and the corresponding mean values of the plateau force for each measured velocity, represented by the dashed black lines. The data were cropped in the x-scale at 1.0 $\mu$m for better visualization.

In Fig. 5 the mean unwinding force vs. extension velocity for all analyzed CS2 fimbriae ($n$ = 30) is summarized. In the graph two distinct regions of structural response to extension velocity can be identified.

For low velocities, less than ~1.0 $\mu$m/s, the force response is independent of extension velocity and the unwinding force is ~10 pN. For extension velocities above ~1.0 $\mu$m/s, the unwinding force increases logarithmically with increasing velocity. This particular transition point is denoted the corner velocity, $\dot{L}^*$ (28). Thus, for velocities below the corner velocity fimbria extension occurs independent of speed and the experiment is performed under steady-state conditions, whereas for velocities above the corner velocity ( $\dot{L}^*$ < $\dot{L}$ ) fimbriae enter a dynamic response region (39).

To find the bond length, the distance from the ground state to the transition barrier, $\Delta x_{AT}$ and the corner velocity $\dot{L}^*$ of CS2 fimbriae and to see if the sticky-chain model (33), a model that describes the behavior of a helix-like polymer under force, could be fitted to the force vs. extension velocity data, we numerically fitted the full set of rate equations, i.e., Eq. (4). The fit is shown by the red dashed line in Fig. 5, with $T$ set to 293 K. The corresponding model parameters were set according to the best fit of the model to our experimental data: $\dot{L}^* = 1300 \pm 200$ nm/s, $\Delta x_{AT} = 0.86 \pm 0.10$ nm, and $\Delta x_{AB} = 5.0 \pm 0.5$ nm.

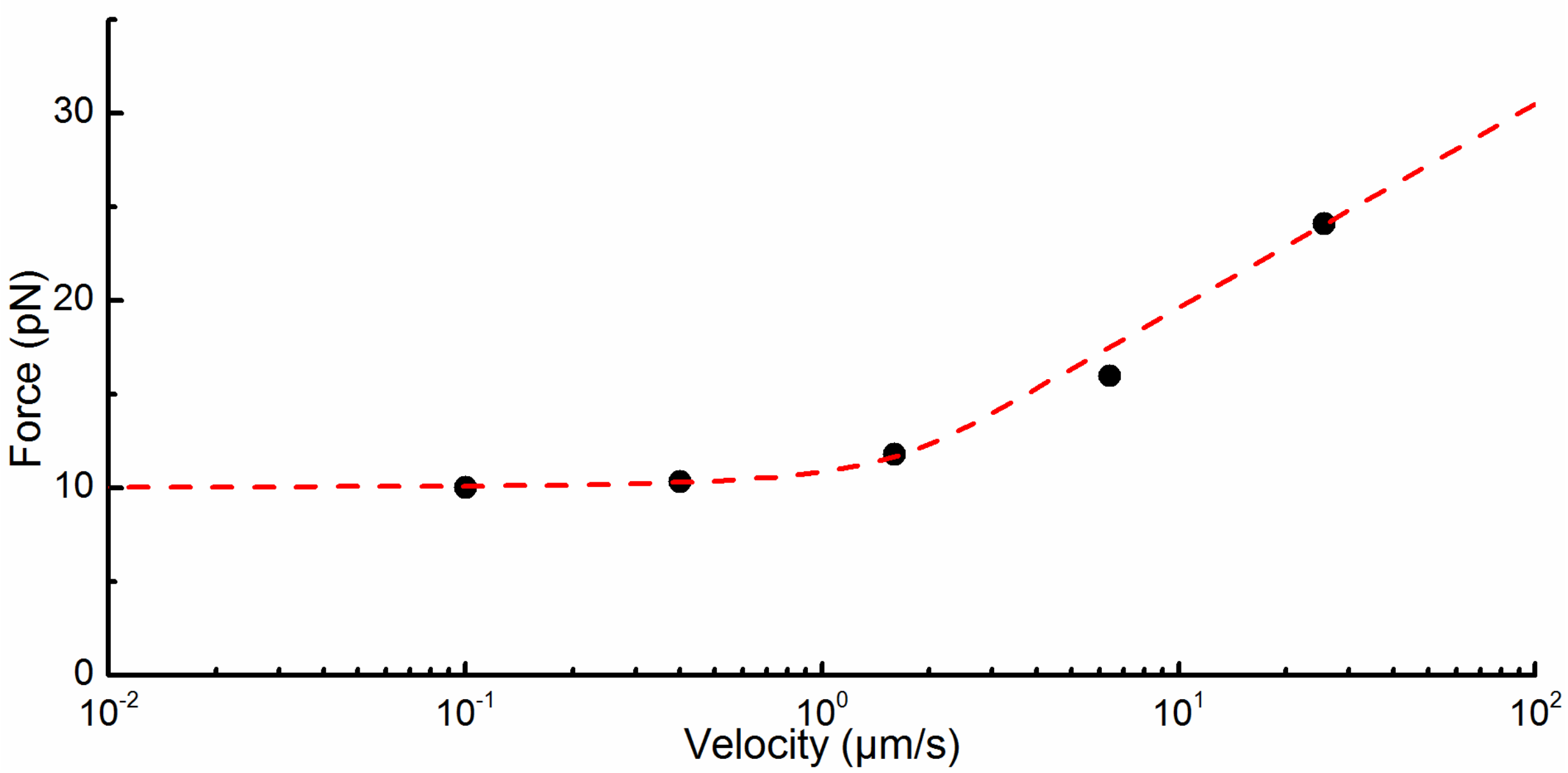

**Figure 5. Dynamic force spectroscopy measurement of CS2 fimbria.** The quaternary structure of an individual CS2 fimbria was unwound at velocities of 0.1, 0.4, 1.6, 6.4, and 25.6 $\mu$m/s for a distance of ~ 2 $\mu$m and the corresponding force responses were sampled at 5 kHz. Each data point (black dots) shows the average unwinding force of fimbriae at five distinct velocities. For low velocities, i.e., velocities below the corner velocity, the unwinding force is independent of speed and amounts to 10 pN. The red dashed line shows a fit of the helix-like polymer model to the data yielding the corner velocity $\dot{L}^* = 1300$ nm/s and the bond length $\Delta x_{AT} = 0.86$ nm, respectively.

## Discussion

Adhesion of pathogenic bacteria to host cells is the initial step of colonization. Adhesion fimbriae are thus key virulence factors, and a detailed understanding of their structural and functional role in bacterial adhesion is essential for elucidating the mechanism of infection. Many fimbriae have the ability to extend and contract when the bacteria are exposed to external forces and it has been shown that unwinding of fimbriae help the bacteria to withstand shear forces by reducing the load on the adhesin protein at the fimbrial tip (22, 40, 41). CFA/I and CS20 fimbriae, expressed by ETEC that colonize the small intestine, are helical and can be unwound to several times their native length. Whether or not CS2 exhibit structural and biophysical similarities with that of CFA/I and CS20, i.e., if CS2 should be classified as helix-like fimbriae that can be unwound and thereby be significantly elongated, was prior to this study not known.

Fimbriae are stabilized by layer-to-layer interactions, which implies that the unwinding force of a fimbriae is related to the number of *ll* interactions (24). Since neither the macromolecular structure of the CS2 fimbriae nor the crystal structure of the major pilin subunit CotA have been solved, we first looked at AFM micrographs to elucidate on the macromolecular structure and to identify any possible morphologies. The expressed CS2 fimbriae were found to be approximately a micrometer in length and were, primarily, in a wider state thus suggesting that they are in a wound state. However, the presence of a few observed narrow width segments suggested that CS2 could make a morphological change from a wide to narrow structure by breaking *ll* interactions formed by subunits in adjacent layers.

To investigate the properties of individual subunits we therefore aligned the major protein CotA using the CFA/I major protein CfaB as a template. Comparison of amino acid composition of the two structures showed 51% identity and 80% similarity. Our subsequent homology model revealed a slightly higher negative surface potential of CotA compare with that of CfaB (23). The negative surface was thereafter confirmed by force spectroscopy measurements with positively charged amidine beads.

Since it has been shown earlier that for some fimbriae niche identity rather than amino acid composition of the major subunit conveys the physical properties of adhesion fimbriae, i.e., the unwinding force of CS20 (15 pN) is more similar to CFA/I (7.5 pN) than P fimbriae (28 pN) even though CS20 share a higher amino acid identity with that of P fimbriae (22.8 % identity) than CFA/I (19.7 % identity) (26), we investigated this correlation using force spectroscopy technique. Force spectroscopy measurements on single fimbriae at steady-state revealed that CS2 is highly flexible and unwinds at a constant force of 10 pN. First, unwinding and rewinding at a constant force suggests that CS2 has a helix-like macromolecular structure similar to both CFA/I and CS20 (33). In addition, the CS2 force response with a - linearly increasing-, constant-, linearly increasing-force, is in line with what has been observed for other helix-like fimbriae such as; P, Type 1, and Type 3 fimbriae (42, 43). This specific force response was also modelled, in a recent work, using a rigid-body model assembled into a helical structure exposed to tensile force (44). Second, the unwinding force level, which is slightly higher than CFA/I and slightly lower than CS20 places CS2 interestingly between these other two ETEC expressed fimbriae colonizing the small intestine.

To analyze if the dynamic properties of CS2 were similar to CFA/I we carried out DFS measurements. The unwinding velocity was increased in steps to measure the unwinding force required for a given velocity. A physical model describing the force response of a helix-like polymer was fitted to the mean unwinding force for each velocity to assess the corner velocity as well as the bond length. The corner velocity provides information of the dynamics of the fimbriae, i.e., fimbriae with a high corner velocity can be unwound at high velocities without responding to an increase in resistance. This implies that a bacterium will "go with the flow" up to the corner velocity of a fimbriae with the unwinding force as the only resisting force. A high corner velocity thereby provides a higher force buffering capability than a low and suggest that fast fluctuating forces are damped out easier. The model fitted the DFS data well, as can be seen in Fig. 5. The corner velocity of CS2 (1300 ± 200 nm/s) is similar to that of CFA/I (1400 ± 200 nm/s). Also the bond length of CS2, derived from DFS measurements, yielded a value of $\Delta x_{AT} = 0.86 \pm 0.1$ nm, which is similar to the bond length of CFA/I, $\Delta x_{AT} = 1.1 \pm 0.1$ nm. From the data presented above we can thus conclude that the similarity of pilin subunits, unwinding forces, corner velocities, and bond lengths strongly indicates that CFA/I and CS2 that are expressed in the same niche are structurally and biophysically similar fimbriae.

This study together with data from the literature suggests that helix-like fimbriae expressed by e.g., pathogenic *E. coli* and *Klebsiella pneumonia,* where the latter is associated with respiratory tract infections, could be categorized into three fimbrial mechanical groups: the low- (~15 pN), the medium- (~30 pN), and the high-force (~60 pN) unwinding fimbriae. ETEC fimbriae such as CS2, CFA/I and CS20 requires <15 pN unwinding force despite differences in their assembly mechanism, see table S1 and refs (24, 26); UPEC and meningitis-associated (MNEC) strains of *E. coli* express fimbriae requiring 21-30 pN of unwinding force (33, 39, 42); however, Type 3 fimbriae expressed by *Klebsiella pneumonia* in the respiratory tract require 65 pN of unwinding force (43). One should also note that the T4 fimbriae expressed by *Streptococcus pneumoniae* that also colonize the respiratory tract are significantly stiffer than UPEC and ETEC expressed fimbriae (45). The analogy between the niche and the biomechanical features of fimbriae has been suggested in earlier studies, and results in this study support that hypothesis by placing CS2 in the low-force unwinding group of fimbriae, where other ETEC expressed fimbriae are also found.

## Author Contributions

N.M carried out the force spectroscopy experiments, DFS measurement and length analysis of fimbriae, B.S prepared the strains, M.A and E.B carried out the homology modeling, M.A and J.Z performed the modelling, and J.Z developed the software used in the force spectroscopy and DFS measurements. All authors contributed to the planning of the studies and to interpretation of the results. N.M, B.S, E.B, and M.A drafted the main manuscript text and all authors reviewed the final version of the manuscript.

## Acknowledgments

This work was supported by NIH (GM05722 and RR025434 to E.B.), the Swedish Research Council (621-2013-5379 to M.A. and the Carl Trygger foundation to M.A. We are grateful to Dr. Stephen Savarino for providing the reagents in this work and Monica Persson for assistance with AFM micrographs.

May 19, 2015

# Supplementary Material - Biomechanical and Structural features of CS2 fimbriae of Enterotoxigenic *Escherichia coli*

Narges Mortezaei[a], Bhupender Singh[a,b], Johan Zakrisson[a], Esther Bullitt[c] and Magnus Andersson[a,b*]

[a]Department of Physics, Umeå University, [b]Umeå Centre for Microbial Research (UCMR), Umeå University SE-901 87 Umeå, Sweden, [c]Department of Physiology and Biophysics, Boston University School of Medicine, Boston, MA 02118, USA

## Mean length of CS2 fimbriae using AFM micrograph data

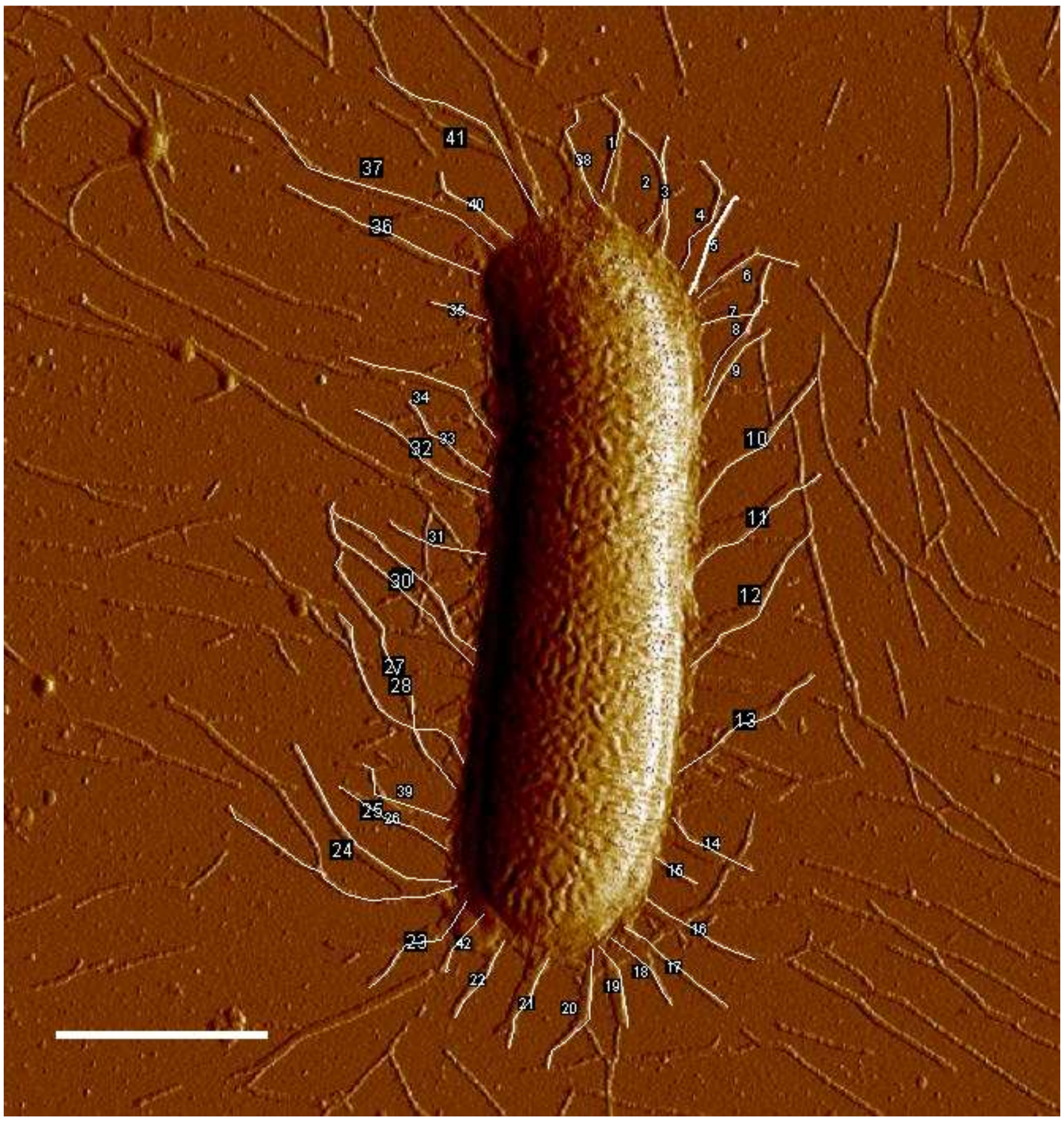

**Figure S1.** Atomic Force Microscopy of C91F cell expressing CS2 fimbriae. The mean length of CS2 fimbriae expressed by C91F cells was assessed by spline fitting to wound sections of fimbriae in AFM micrographs. The mean length was found to be ~ 0.88 ± 0.34 $\mu$m ($n$ =270). Scale bar is 1.0 $\mu$m.

May 19, 2015

# Sequence alignment of CotA and CfaB

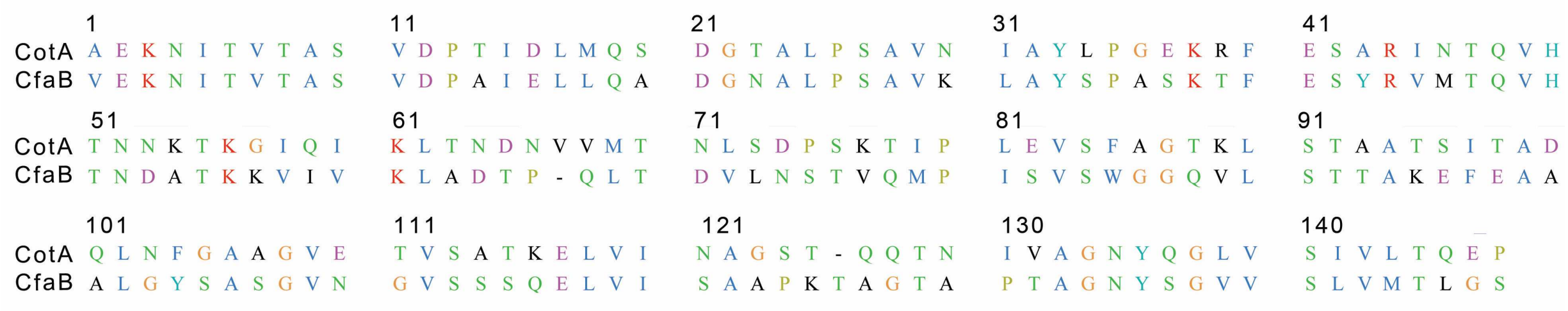

**Figure S2.** Sequence alignment of the CotA and CfaB, the major subunits of the CS2 and CFA/I fimbriae, respectively. Numbering corresponds to the CotA sequence.

May 19, 2015

## Side-by-side comparison of the surface properties of CfaB and CotA

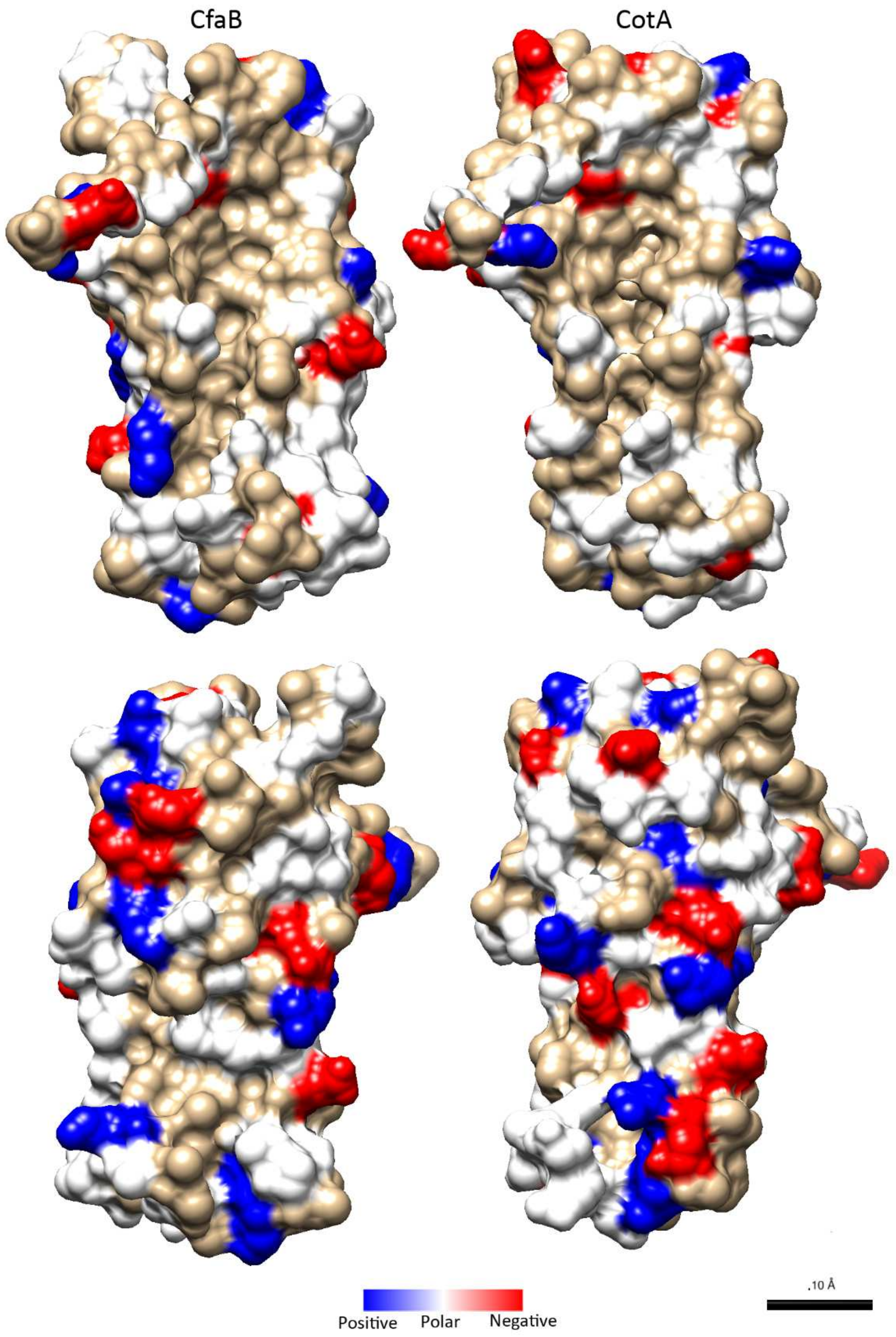

**Figure S3.** Surface views of the charge distribution of CfaB (left) and CotA (right) subunits. The upper panel represent the inner surface, whereas the lower panel represents the outer surface of the subunits. Scale bar is 10 Å.

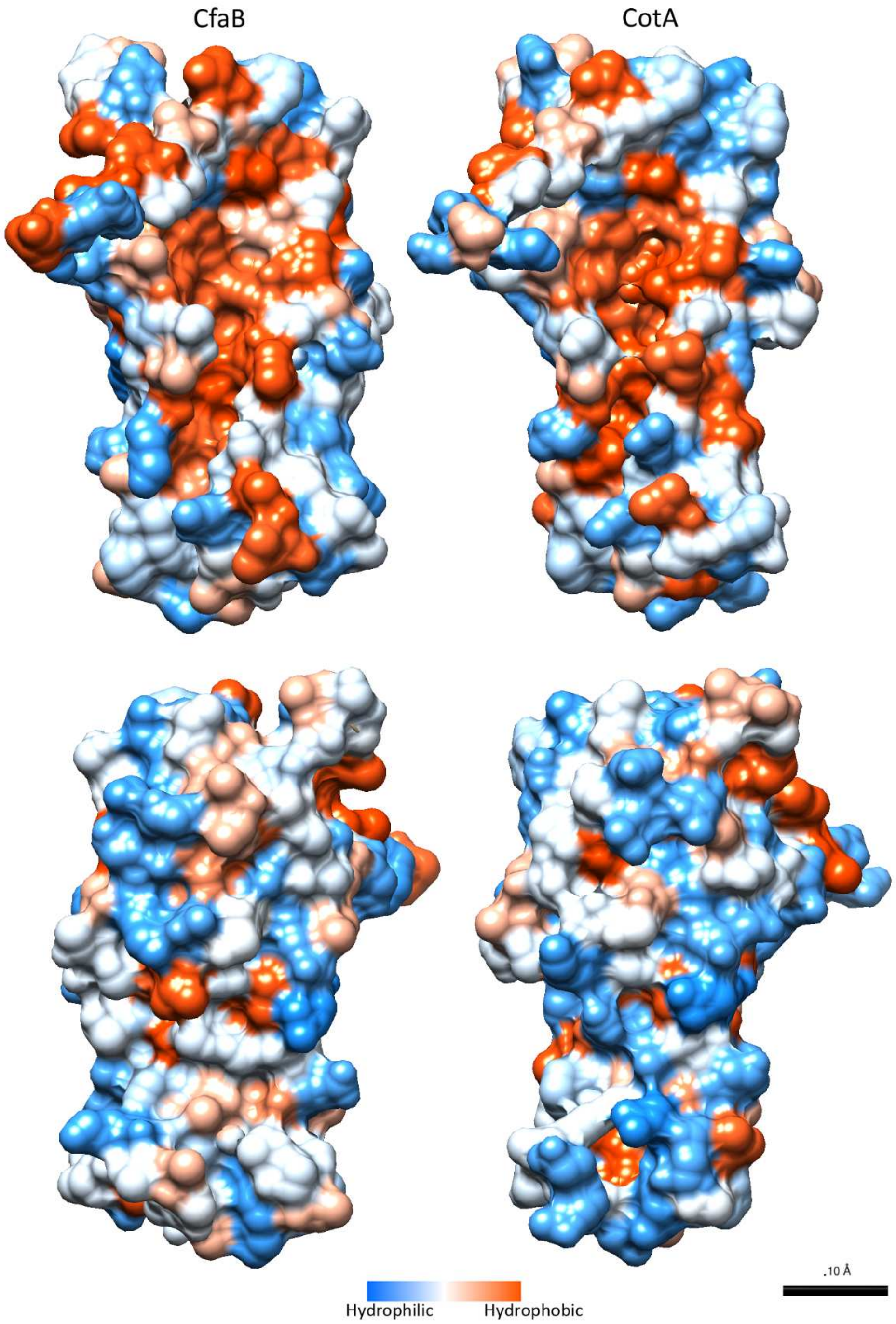

**Figure S4.** Surface views of the hydrophobicty and hydrophilicity of CfaB (left) and CotA (right) subunits. The upper panel represents the inner surface, whereas the lower panel represents the outer surface of the subunits. Scale bar is 10 Å.

## Consecutive force spectroscopy measurement of CS2 fimbriae

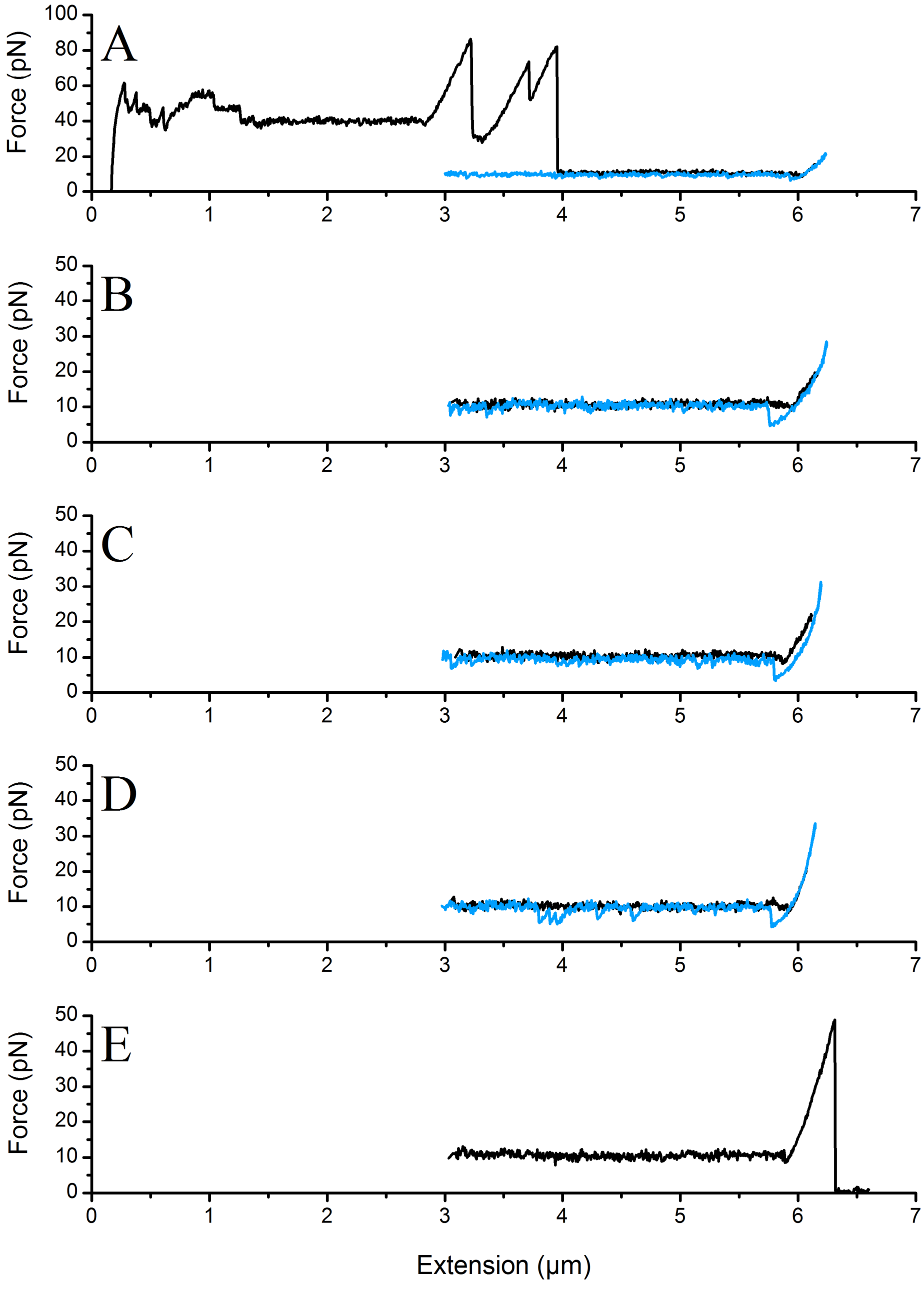

**Figure S5.** Consecutive force measurement of a CS2 fimbria. Panels A to E shows sequential unwinding (black) and rewinding (blue) of a fimbria. In panel A, several fimbriae are initially attached to the probe bead giving rise to a higher unwinding force. During extension the shortest fimbriae detached and only one fimbria remained attached. The fimbria were then rewinded, blue curve in panel A, and thereafter several consecutive unwinding/rewinding cycles were performed, see panels B-D. In panel E the fimbria detaches from probe bead, allowing us to measure any offset drifts that occurred during the measurements.

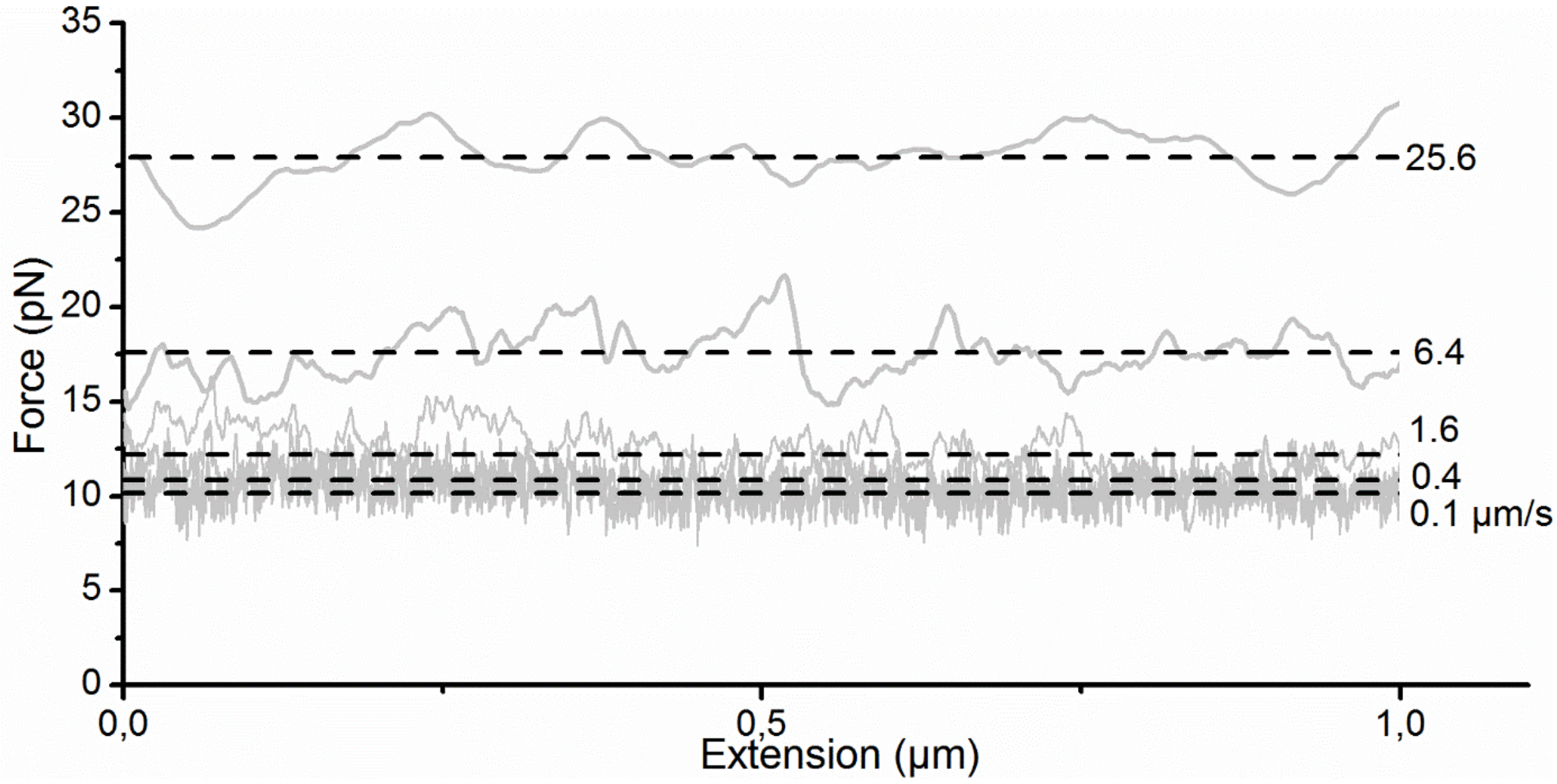

**Figure S6.** A DFS measurement of a CS2 fimbria at five given velocities. The grey curves show the force vs. extension data of a CS2 fimbria at the given velocities (0.1, 0.4, 1.6, 6.4, 25.6 $\mu$m/s). The black dashed lines represent the corresponding mean value of the force plateaus at the given velocities. For better visualization the force data are shown up to 1.0 $\mu$m. Detailed description of dynamic force spectroscopy theory on bacterial fimbriae can be found in the following references (1–3).

May 19, 2015

## Supplementary Tables

**Table S1. Parameter values of CS2, CFA/I, and CS20 fimbriae.** A comparison of the unwinding force, the bond length, and the corner velocity for CS2, CFA/I and CS20 fimbriae.

| **Strain/plasmid** | **C91F** | **HMG11/pNTP119**(4) | **WS7179A2/pRA101**(5) |
|---|---|---|---|
| **Fimbriae** | **CS2 (this work)** | **CFA/I** | **CS20** |
| $F_{uf}$ $(pN)$ | 10 ± 1.5 | 7.5 ± 1.5 | 15 ± 1 |
| $\Delta x_{AT}$ $(nm)$ | 0.86 ± 0.10 | 1.1 ± 0.1 | 0.4 ± 0.09 |
| $\dot{L}^*$ $(nm/s)$ | 1300 ± 200 | 1400 | 877 |